\documentclass[prb, reprint, superscriptaddress]{revtex4-2}

\usepackage{graphicx}
\usepackage{dcolumn}
\usepackage{bm}
\usepackage{xcolor}
\begin{document}


\title{Metastable Charge Distribution Between Degenerate Landau Levels}

\author{Wenlu Lin} 
\affiliation{International Center for Quantum Materials, 
  Peking University, Haidian, Beijing 100871, China}
\author{Xing Fan} 
\affiliation{National Laboratory of Solid State Microstructures $\&$ Department of Materials Science and Engineering, College of Engineering and Applied Sciences, Nanjing University, Nanjing 210093, China}
\author{Lili Zhao} 
\affiliation{International Center for Quantum Materials, 
Peking University, Haidian, Beijing 100871, China} 
\author{Yoon Jang Chung} 
\affiliation{Department of Electrical Engineering, 
Princeton University, Princeton, New Jersey 08544, USA} 
\author{Adbhut Gupta} 
\affiliation{Department of Electrical Engineering, 
Princeton University, Princeton, New Jersey 08544, USA}
\author{Kirk W. Baldwin} 
\affiliation{Department of Electrical Engineering, 
Princeton University, Princeton, New Jersey 08544, USA}
\author{Loren Pfeiffer} 
\affiliation{Department of Electrical Engineering, 
Princeton University, Princeton, New Jersey 08544, USA}
\author{Hong Lu} 
\affiliation{National Laboratory of Solid State Microstructures $\&$ Department of Materials Science and Engineering, College of Engineering and Applied Sciences, Nanjing University, Nanjing 210093, China}
\affiliation{Shishan Laboratory, Suzhou Campus of Nanjing University, Suzhou 215000, China}
\author{Yang Liu} 
\email{liuyang02@pku.edu.cn} 
\affiliation{International Center for Quantum Materials, 
  Peking University, Haidian, Beijing 100871, China} 

\date{\today}

\begin{abstract}

  We study two dimensional electron systems confined in wide quantum
  wells whose subband separation is comparable with the Zeeman
  energy. Two $N=0$ Landau levels from different subbands and with
  opposite spins are pinned in energy when they cross each other and
  electrons can freely transfer between them. When the disorder is
  strong, we observe clear hysteresis in our data corresponding to
  instability of the electron distribution in the two crossing
  levels. When the intra-layer interaction dominates, multiple minima
  appear when a Landau level is $\frac{1}{3}$ or $\frac{2}{3}$ filled
  and fractional quantum hall effect can be stabilized.
  
\end{abstract}

\maketitle

\section{INTRODUCTION}

The quantum Hall effect is an incompressible quantum liquid phase
signaled by the vanishing of the longitudinal conductance and the
quantization of the Hall conductance seen in two-dimensional electron
systems (2DES) at large perpendicular magnetic field \cite{Tsui.PRL.1982, vonKlitzing.IQHE.Novel}. 2DES with spin and subband degree of freedom has additional sets of Landau levels (LLs)
separated by the Zeeman energy $E_Z$ or subband separation
$\Delta_{SAS}$, respectively \cite{Muraki.PRB.2008, Liu.PRB.2011,
  ZhangDing.PRB.Bilayer,P.Kim.InterlayerCharges, LiJIA.Mat.Phy.2019,
  ShiQH.Bilayer.NatureNano}. When only a small number of LLs are
occupied at LL filling factors $\nu<4$, the electron distribution
between the two subbands deviates from the $B=0$ subband
densities. The 2DES's Hartree potential leads to the renormalization
of the subbands' energies and wavefunctions so that the total energy
is minimized. When two LLs are degenerate at the Fermi energy, the
electrons can redistribute between them with negligible energy cost \cite{ Trott.PRB.1989, Liu.PRL.2011B, Nubler.5/2}.

 \begin{figure}
 \includegraphics[scale=0.95]{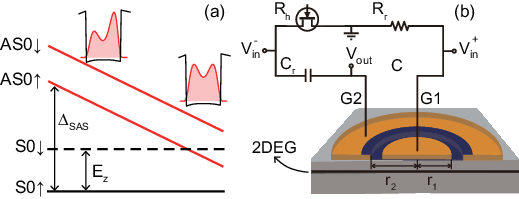}%
 \caption{\label{fig0}(a) Relevant energy levels and charge distribution in wide quantum well. (b) Schematic of gates with Corbino-like geometry for capacitance measurements.}
\end{figure}

 \begin{figure*}
 \includegraphics[scale=1]{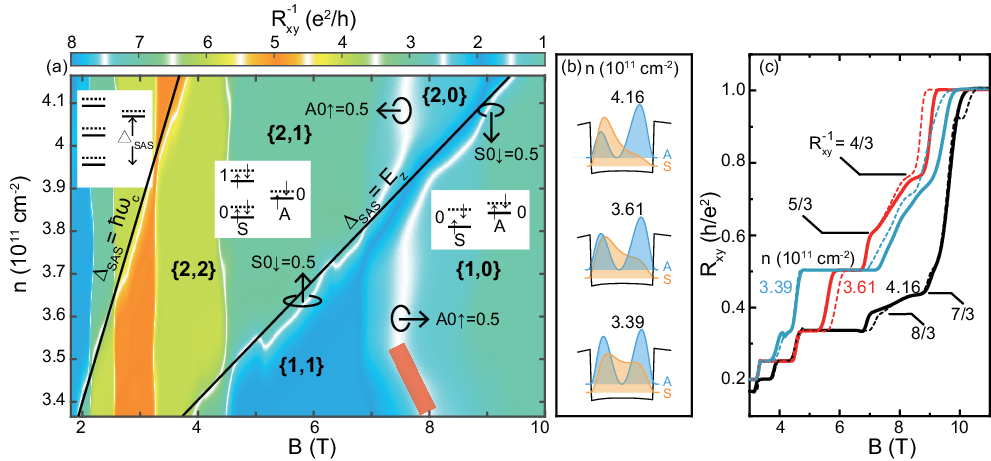}%
 \caption{\label{fig1}(a) Color-coded plot of $R^{-1}_{xy}$ as a
   function of $n$ and $B$. The insets are LL diagrams corresponding
   to three regions separated by the $\Delta_{SAS}=\hbar\omega_C$ and
   $\Delta_{SAS}=E_z$ lines. Plaques with different colors correspond to different
   $R_{xy}$ plateaus, where \{$\nu_{S}$, $\nu_{A}$\} labels the number
   of LLs belong to the two subbands whose electron filling factor is
   higher than 1/2. (b) Calculated charge distribution ($|\psi_S|^2$ and
   $|\psi_A|^2$) for the two subbands of 45-nm-wide quantum well sample at different n. (c) $R_{xy}$ vs. $B$ traces at different $n$. We use solid and dashed traces for up- and down-sweep respectively in all figures throughout this manuscript.}
\end{figure*}

 \begin{figure}
 \includegraphics[scale=0.95]{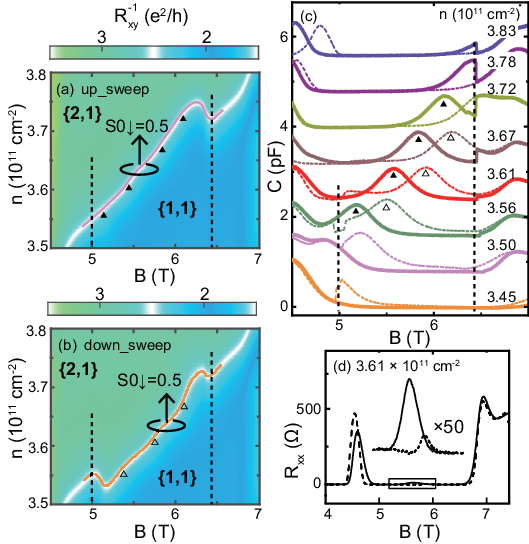}%
 \caption{\label{fig2}(a, b) Color-coded plot of $R^{-1}_{xy}$ as a function of $n$ and $B$ near the \{1,1\}-to-\{2,1\} boundary of the up- and down-sweep data respectively. The pink and orange lines highlight the positions of plateau-to-plateau transition where clear hysteresis appears. (c) $C$ vs $B$ traces at different $n$. We mark the $C$ peak in the up- and down-sweep data and summarize their positions in panel (a, b) with solid and open symbols, respectively. (d) $R_{xx}$ vs. $B$ trace of density 3.61, and inset shows hysteresis after 50 times magnification. }
\end{figure}

High sensitivity, local capacitance measurement reveals fine
structures and delicate quantum phases of the 2DES \cite{Mosser.SSC.1986, Smith.PRB.1986, Ashoori.PRL.1992,
  Eisenstein.PRL.1992, Zibrov.nature.2017, Irie.APE.2019,
  ShiQH.EvenAndOddFQHE, Deng.prl.2019, Tomarken.PRL.2019}. A recent
examination by Lili Zhao \emph{el al.} finds that the capacitance $C$
and the conductance $\sigma$ are intertwined and both of them reflect
the 2DES's transport properties \cite{Zhao.CPL.2022}. In this work, we
study the 2DES confined in wide quantum wells using the same technique
as ref. \onlinecite{Zhao.CPL.2022} and discover features such as
hysteresis and splitting minima, consistent with the pinning of two
LLs and metastable charge distribution \cite{Pollanen.SAW.ChargeInstability, Tutuc.layercharge.instability.transport, Deng.PRB.2017}.

\section{\uppercase{SAMPLES and methods}}

We study two GaAs/AlGaAs samples grown by molecular beam epitaxy where the 2DESs are confined in the wide quantum wells and electrons occupy two subbands, leading to additional subband degree of freedom. We use back gate to change density and $\Delta_{SAS}$ increases when the charge distribution becomes imbalanced, see in Fig. 1(a). We study samples from two different wafers grown by different groups. The 45-nm-wide quantum well sample has 4.0 \footnote{We omit the unit of the charge density 10$^{11}$ cm$^{-2}$ and the unit of the conductance (resistance) $e^2/h$ ($h/e^2$) throughout this manuscript for brevity.} as grown density and $6\times 10^{4}$ cm$^{2}$/(V$\cdot$s) low temperature mobility and the 80-nm-wide quantum well sample has 1.1  as grown density and $6\times 10^{6}$ cm$^{2}$/(V$\cdot$s) low temperature mobility. The $\Delta_{SAS}$ of the low mobility sample can be measured from the LL crossings seen when $\Delta_{SAS}$ = $\hbar \omega_{C}$ and $\Delta_{SAS}$ = $E_{Z}$, and the $\Delta_{SAS}$ of the high mobility sample can be extracted from the Fourier transform of the low field Shubnikov-de Hass oscillations. In both samples, $\Delta_{SAS}$ is comparable with the exchange-enhanced $E_Z$ so that the
S0$\downarrow$ and A0$\uparrow$ levels are close in energy (S and A
refer to symmetric and antisymmetric subbands, 0 refers to the $N=0$
LL index, $\uparrow$ and $\downarrow$ refer to up- and down-spin) \cite{Muraki.PRL.2001, Zhang.PRB.2006, Liu.PRB.2015}. Although the subband wavefunctions are not strictly symmetric or antisymmetric when charge distribution is imbalanced, for simplicity, we still use S and AS for the two subbands.

Each sample is a 2 $\times$ 2 mm$^{2}$ cleaved piece, with eight alloyed InSn
contacts; see the Fig. 1(b). We fit the samples with an In back gate to tune the electron density, and evaporate multiple Ti/Au front gates with Corbino-like geometry for capacitance measurement. The outer radius of G1 and the inner radius of G2 are r1 = 120 $\mu$m and r2 = 140 $\mu$m, respectively; see the Fig. 1(b). Capacitance is measured between the two gates using a cryogenic bridge and excitation frequency up to 140 MHz \cite{Zhao.RSI.2022}. The capacitance($C$) and conductance($G$) components can be extracted from the in-phase and out-phase of the bridge output $V_{out}$ \cite{Zhao.PRL.2023}. The longitudinal and Hall resistances, $R_{xx}$ and $R_{xy}$, can be measured \emph{in-situ} using standard lock-in technique($<$40 Hz). All the experiments are performed in a dilution refrigerator whose base temperature is 10 mK.

\section{RESULTS AND DISCUSSION}

Fig. 2(a) shows the color-coded plot of $R^{-1}_{xy}$ as a function
of magnetic field $B$ and electron density $n$, taken from the 45-nm-wide quantum well sample. In Fig. 2(a), colored plaques represent different quantized Hall conductances $\sigma_{xy} = R^{-1}_{xy}$ where $R_{xx}$ = 0, and plateau-to-plateau transitions appears as white ribbons. This sample has low mobility so that the disorder dominates and each Landau level contributes zero/one Hall conductance \cite{Note1} if its filling factor is below/above 1/2, respectively. The plateau-to-plateau transitions appear when a LL is exactly half filled.

We highlight two series of transitions by the two black lines which correspond to the conditions $\Delta_{SAS}=\hbar\omega_c$ and $\Delta_{SAS}=E_Z$ as labeled in Fig. 2(a) \cite{Liu.PRB.2011,
  ZnO.Falson.Nat.Phy}. The charge distribution is balanced when $n \simeq$
3.0 and the higher density leads to a more imbalanced quantum well \cite{Note1}. $\Delta_{SAS}$ increases from about 40 to 80 K as $n$
increases from 3.4 to 4.1, deduced from the
$\Delta_{SAS}=\hbar\omega_c$ line \cite{Liu.PRB.2011}. The
$\Delta_{SAS}=E_Z$ line appears at about twice the magnetic field than
the $\Delta_{SAS}=\hbar\omega_c$ line, suggesting that $E_z$ is
significantly enhanced by about 30 times because of the exchange energy \cite{Muraki.PRL.2001,
  Zhang.PRB.2006, Liu.PRB.2015}. These two lines separate Fig. 2(a)
into three zones with different LL configurations shown by the
insets. We label the number of more-than-half-filled LLs from the
symmetric and antisymmetric subbands as \{$\nu _{S}$ , $\nu _{A}$\} for
each colored plaque in Fig. 2(a) \cite{Liu.PRB.2011, ShiQH.Bilayer.NatureNano, ZnO.Falson.Nat.Phy}.

It is quite surprising that the \{1,0\}-to-\{1,1\} boundary, at which
the A0$\uparrow$ level is half-filled, moves towards lower field while
the total density increases, see the thick red line. This anomalous
phenomenon can be better illustrated by Fig. 2(c), where the $R_{xy}$
of the $n=3.61$ trace is larger than the $n=3.39$ trace at
$B\gtrsim 6$ T. The kink at $R_{xy}=3/5$ and 3/4 \cite{Note1} in the
$n=3.61$ trace signals the formation of fractional quantum Hall effect
when the A0$\uparrow$ is 1/3 and 2/3 filled, respectively. According
to the total filling factor, the S0$\downarrow$ level is
less-than-half filled at about 0.4 and contributes zero Hall
conductance. On the \{2,0\}-to-\{2,1\} boundary, the $R_{xy}$ trace at $n=4.16$ exhibits kinks at $R_{xy}=3/7$ and 3/8 at similar $B$, signaling that the A0$\uparrow$ is 1/3 and 2/3 filled when the S0$\downarrow$ level is more-than-half filled at about 0.7 and contributes $e^2/h$ in Hall conductance. The above observations suggest that the S0$\downarrow$ and A0$\uparrow$ levels are both partially filled and energetically
degenerate at the Fermi energy in a large range of $B$ and $n$, i.e. the two crossing LLs are pinned together and both of them will be partially occupied in a finite range of $B$ and $n$
\cite{Trott.PRB.1989, Liu.PRL.2011B, Nubler.5/2, CHW.PRB.1999, SJGeier.PRB.2002}.

\begin{figure*}[htbp]
 \includegraphics[scale=0.95]{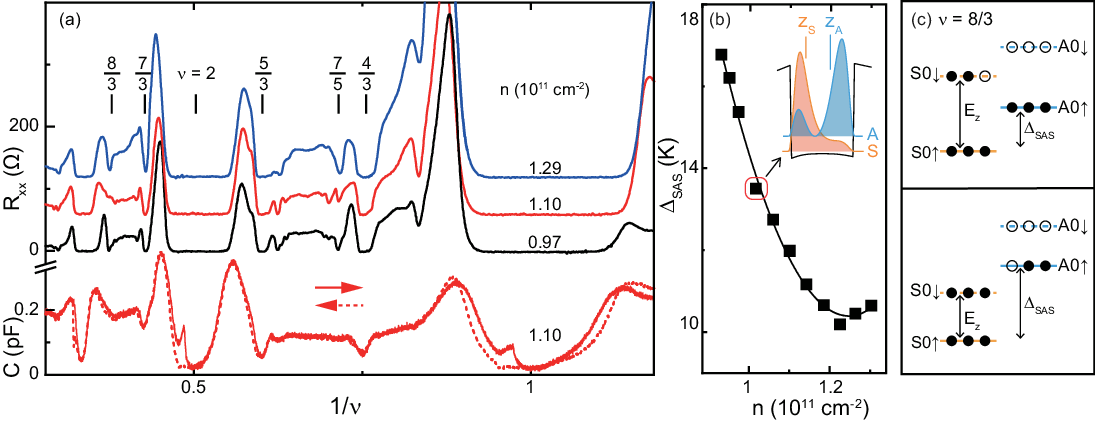}
 \caption{\label{fig3}(a) The longitudinal resistance ($R_{xx}$) and
   capacitance ($C$) taken from the 80-nm-wide quantum well
   sample. Hysteresis of $C$ can be observed within the
   $R_{xx}$ plateau region at integer quantum Hall effects. (b) The experimentally measured low field
   $\Delta_{SAS}$ deduced from the Shubnikov–de Haas oscillation. We
   also show calculated charge distribution ($|\psi_S|^2$ and
   $|\psi_A|^2$) for the two subbands, and the vertical bars label $z_{S}$ and $z_{A}$ mark their centroid
   $<\psi|\hat{z}|\psi>$. (c) Diagram shows the two configurations for stable $\nu$ = 8/3 fractional quantum Hall states in two-subband picture. The electrons and holes are marked as solid and open circles, respectively. We use red and blue lines to distinguish symmetric and antisymmetric subband.}
\end{figure*}

Hysteresis occuring at the \{1,1\}-to-\{2,1\} boundary is shown in Fig. 3(a) and 3(b), across which the filling factor of the S0$\downarrow$ level varies through 1/2, see also the $n=3.61$ data near 6 T in Fig. 2(c). This hysteresis can be better seen in Fig. 3(c). The capacitance $C$
vanishes when the system forms a quantum Hall effect and a peak
appears whenever a LL is exactly half-filled
\cite{Zhao.CPL.2022}. More specifically, the peaks in the up- and
down-sweep traces, where the S0$\downarrow$ level is half filled,
appear at different fields. We mark the capacitance peak seen in the
up- and down-sweep traces and summarize their positions in Fig. 3(a) and Fig. 3(b)
with solid and open symbols, respectively. The hysteresis, i.e. the
peak position depends on the sweep direction, indicating a metastable
charge distribution in the two energetically locked LLs. The sharp
capacitance jumps near the dashed lines mark the onset of the
discrepancy between the up- and down-sweeps, possibly due to the
completely emptying or filling of the A0$\uparrow$
level \cite{Zhu.Charge.ImpurityLayer}. We also show transport results in Fig. 3(d) at $n=3.61$ where a barely visible $R_{xx}$ peak appears at the \{1,1\}-to-\{2,1\} boundary and a hysteresis behavior is also seen, consistent with previous reports \cite{DePoortere.PRL.2003, Pan.PRB.2005, Vincenzo.NATURE.1999}. The $R_{xx}$ peak is a narrow spike appearing only at the boundary when one LL is exactly half-filled, while the hysteresis region in our capacitance data is much broader with two sharp ends far from the boundary, see the two dashed lines in Fig. 3(d). This is because our capacitance measurement is a local probe achieved by gates without contacts involved, so that it is sensitive to local domains and charge instability caused by LL pinning and domains. The transport measurement, on the other hand, averages across the whole sample and the domains can be seen only when they form a percolation network at the plateau-to-plateau transition boundary. Therefore, we can observe hysteretic behavior in C measurements while the transport features are barely visible.

The charge transfer is universally present in imbalanced 2DES confined in wide and double quantum wells \cite{Liu.PRL.2011B, Nubler.5/2}. Empirically, visible hysteresis phenomena are usually seen strong when the 2DES forms strong insulating phases so that the time constant for charge transfer is sufficiently long \cite{Tutuc.layercharge.instability.transport, Deng.PRB.2017}. It is also seen if domain structure forms when two LLs with different spins cross \cite{DePoortere.PRL.2003, Pan.PRB.2005, Vincenzo.NATURE.1999}. In general, the random disorder plays an essential role in the hysteresis. We
study the low-disorder 80-nm-wide quantum well sample below. The
low-field subband separation $\Delta_{SAS}$ measured from the Fourier
transform of the Shubnikov-de Haas oscillations is shown in the
Fig. 4(b). It increases from 10 K to about 17 K when we reduce $n$ from
1.3 to 0.88 by back-gate voltage. In Fig. 4(a), the $R_{xx}$ minimum at $\nu=8/3$ disappears in the $n$=1.1 trace signaling that $\Delta_{SAS}$ equals $E_z$ which is $\sim$23-times enhanced
\cite{Liu.PRB.2015}. In the high mobility samples, disorder is reduced so that the electron-electron interaction stabilizes fractional quantum Hall states. Fractional quantum Hall effects can be understood as integer quantum Hall effects of composite fermion which forms by attaching 2 flux quanta to each particle \cite{Jain.CF.2007}. Fractional quantum Hall state at $\nu$ = 8/3 is $\nu$ = 1/3 of holes and is stable when either $\Delta_{SAS} > E_{Z}$ or $\Delta_{SAS} < E_{Z}$; see in Fig. 4(c). It disappears if and only if S0$\downarrow$ and A0$\uparrow$ are degenerate and the disappearance of minima at $\nu$ = 8/3 signals the exact condition $\Delta_{SAS}$ = $E_{Z}$ \cite{Liu.PRB.2015}. Similarly, a weakening of the $\nu=4/3$ and 7/5 states is also visible, suggesting that the S0$\downarrow$ and A0$\uparrow$
levels are close in energy within the $B$ and $n$ range of our
study. In the capacitance trace, hysteresis is only seen near the
integer filling factors $\nu=1$ and 2, where the 2DES consists of
incompressible quantum Hall effect and randomly pinned
quasiparticles/quasiholes \cite{Zhu.Charge.ImpurityLayer}. This hysteresis gradually disappears at high temperatures above 400 mK when the thermal fluctuation softens the disorder pinning, see Fig. 5(a) and (b). It's also quite possible that these dilute quasiparticles/quasiholes may form a Wigner crystal which has a large capacitance response \cite{Zhao.PRL.2023}. In Fig. 5(c), we show $R_{xx}$ at base temperature and 208 mK where no hysteresis at any temperature is seen. This is consistent with Fig. 3 results: $C$ is sensitive to local domain while $R_{xx}$ is an average over the entire sample, so that $C$ can reflect local charge transfer between subbands.

When the partial filling factor in the LLs increases, the electron
interaction stabilizes fractional quantum Hall effects and the
hysteresis disappears. We can deduce the local conductance $G$ between
the two Corbino-like gates \cite{Zhao.CPL.2022}, see Fig. 6(a). The two subbands can be understood as parallel channels. $R_{xx}$ shows minima only if two subbands are both insulating, so that the $R_{xx}$ minima only appear at specific $\nu=$ 7/3 and 8/3, etc. The measured $C$ is the sum of these two parallel channels so that minima can be observed whenever one LL has $\frac{1}{3}$ or $\frac{2}{3}$ insulating states in Fig. 6(a). At $n$=1.1, no minima is seen for the $\nu=8/3$ state when the two configurations
$(\nu_{\text{S0}\downarrow},\nu_{\text{A0}\uparrow})=$
($\frac{2}{3}$,1) and (1, $\frac{2}{3}$) have the same energy, which is consistent with $R_{xx}$ results in Fig. 4(a); $\nu_{\text{S0}\downarrow}$ and $\nu_{\text{A0}\uparrow}$ are the
corresponding filling factor of the S0$\downarrow$ and A0$\uparrow$
levels \cite{Liu.PRB.2015}. $\Delta_{SAS}$ increases as $n$ decreases
so that the S0$\downarrow$ level becomes more populated and the (1,
$\frac{2}{3}$) configuration is energetically favorable. However, the
subbands of asymmetric quantum well has different $z_{S}$ and $z_{A}$,
the center of their charge distribution $<\psi|\hat{z}|\psi>$, which
leads to finite energy cost for redistributing electrons between them,
see Fig. 4(b). Therefore the in-plane Coulomb interaction can no
longer stabilize the 8/3 state at exact total filling factor. Instead,
a minimum reappears but shifts to smaller filling factors,
corresponding to (1-$\nu^*$, $\frac{2}{3}$) where some quasiholes
($\nu^{*}$) appear in the S0$\downarrow$ level. Meanwhile, the 7/3
state has the configuration ($\frac{2}{3}$, $\frac{2}{3}$) when the
S0$\downarrow$ and A0$\uparrow$ levels are degenerate at $n$=1.26 so
that a single minimum appears at exactly 7/3 total filling
factor. When the energy of A0$\uparrow$ level increases at $n \leq$
1.26, this minimum splits into two at which the system has
configuration of ($\frac{2}{3}$+$\nu^*$, $\frac{2}{3}$) and
($\frac{2}{3}$, $\frac{2}{3}$-$\nu^*$), respectively \footnote{We
  would like to emphasize that in quasi-DC transport we always observe
  minimum at proper total filling factors in Fig. 4(a). A more
  comprehensive study is necessary to explain the discrepancy.}.

\begin{figure}
 \includegraphics[scale=0.95]{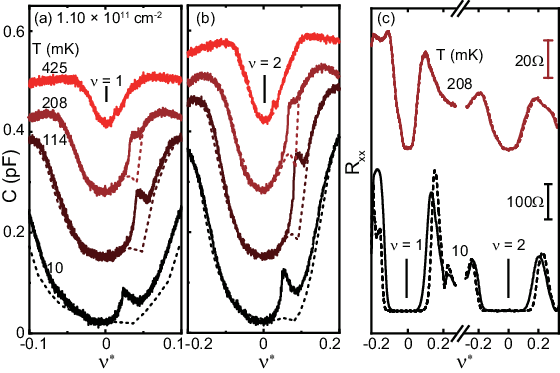}%
 \caption{\label{fig4}(a, b) The hysteresis of $C$ around $\nu$ = 1 and 2 gradually disappears as temperature increases to $\gtrsim$ 400 mK. (c) $R_{xx}$ results at base temperature and 208 mK.}
\end{figure}

\begin{figure*}
 \includegraphics[scale=0.95]{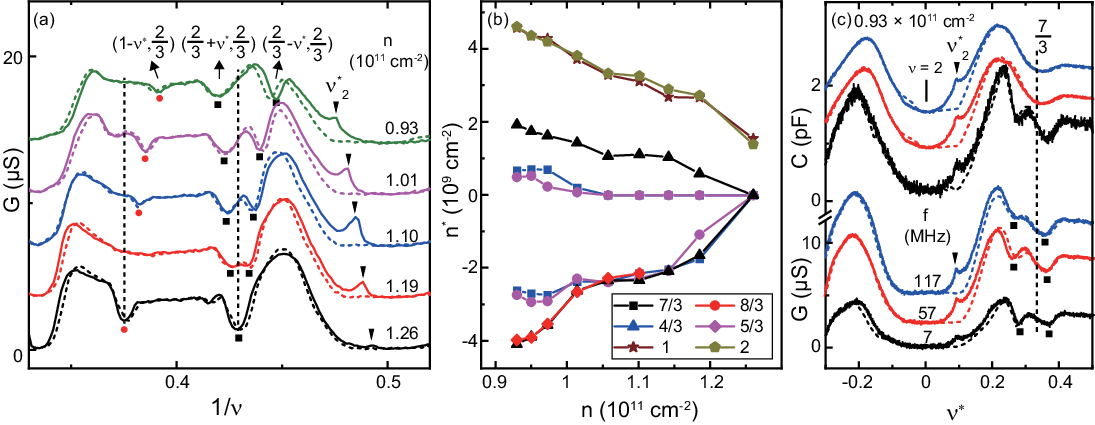}%
 \caption{\label{fig5}(a) Local conductance component extracted from
   the capacitance measurements. Data taken from the 80-nm-wide
   quantum well sample at different densities. We mark hysteresis and
   double minima with different symbols to illustrate their evolution
   as a function of quantum well symmetry. ($\nu_{S0\downarrow}$,$\nu_{A0\uparrow}$) signals LL configuration, and we also use $\nu^{*}$ to mark the filling location of the maximal hysteresis of integer quantum Hall effect. (b) Summary of quasi-particle and quasi-hole densities of different features. (c) Local conductance and capacitance measured by using different frequencies.}
\end{figure*}

We summarize the quasi-particle or quasi-hole densities,
$n^{*} = n \cdot \nu^{*}/\nu$, for several observed $G$ minima in
Fig. 6(b). We note that the hysteresis near $\nu=2$ in Fig. 6(a) also
has a density dependent evolution, where the peak in the up-sweep
gradually shifts to higher filling as $n$ decreases. It is likely that
this peak signals the formation of the integer quantum Hall effect by
spontaneously fill up the A0$\uparrow$ level
\cite{Zhu.Charge.ImpurityLayer}. The residual quasiparticle density
$n^*$ in the S0$\downarrow$ level increases when the density
decreases, which generates a Hartree potential to compensate the
increasing $\Delta_{SAS}$. For example, $n^{*}=0.02$ reduces the
antisymmetric subband energy by
$\simeq e^2 n^{*}/\varepsilon|z_{S}-z_{A}|\simeq 10$ K, where $e$ is
the electron charge, $\varepsilon$ is the GaAs permitivity and
$|z_{S}-z_{A}| \simeq$ 31 nm is the equivalent charge separation
between the two subbands (see Fig. 4(b)). We also summarize the
density of residual charges near $\nu$=1 and 2 in Fig. 6(b).

Fig. 6(c) studies the split minima near 7/3 and the hysteresis near
$\nu=2$ as a function of measurement frequency $f$. The hysteresis
around $\nu=2$ becomes more profound at higher frequencies both in $C$
and $G$, consistent with the assumption that it's the response of
pinned charges. On the other hand, the split minima around 7/3 is a
low-frequency phenomenon, i.e. the $G$ minima becomes shallower at
high frequencies, and only one shallow minimum at 7/3 appears in the
capacitance at $f=57$ and 117 MHz. This may because that the dilute
extra quasiparticles/quasiholes can response to the probing high
frequency electric field and screen the minima. 

\section{\uppercase{conclusion}}

In summary, we closely examine the 2DES confined in wide quantum wells
using capacitance measurement. Our results indicate that the
S0$\downarrow$ and A0$\uparrow$ levels are pinned in energy if the
subband separation and exchange-enhanced Zeeman energy are comparable. We
discover hysteresis and charge instability when the disorder is
strong, and multiple minima at fractional quantum Hall effect if the
intra-layer interaction dominates. Our observations shed light on the
complex internal structure of 2DES when multiple degrees of freedom
present.

\begin{acknowledgments}
  
  We acknowledge support by the National Natural Science Foundation of China (Grant No. 92065104 and 12074010) and the National Basic Research Program of China (Grant No. 2019YFA0308403) for sample fabrication and measurement. This research is funded in part by the Gordon and Betty Moore Foundation’s EPiQS Initiative, Grant GBMF9615 to L. N. Pfeiffer, and by the National Science Foundation MRSEC grant DMR 2011750 to Princeton University.  We thank L. W. Engel, Bo Yang and Xi Lin for valuable discussion.

\end{acknowledgments}

\bibliography{./bib_full_20230729}

\end{document}